\documentclass{research}
\usepackage{standalone}
\usepackage{color}
\usepackage{epsfig,amssymb,amsmath,amsthm,amsfonts,amsbsy,mathrsfs}
\usepackage{graphics}
\usepackage{graphicx,color}
\usepackage{amsmath}
\usepackage{amssymb}
\usepackage[version=3]{mhchem} 
\usepackage{float}
\usepackage{lineno}

\newcommand*\chem[1]{\ensuremath{\mathrm{#1}}}

\title{High-throughput computational screening for bipolar magnetic semiconductors}

\author{Haidi Wang$^{1}$, Qingqing Feng$^{2}$, Xingxing Li$^{2*}$ \& Jinlong Yang$^{2*}$}

\begin{document}

\maketitle

\begin{affiliations}
\item School of Physics, Hefei University of Technology, Hefei, Anhui 230601, China
 \item Hefei National Laboratory for Physical Sciences at Microscale, Department of Chemical Physics, and Synergetic Innovation Center of Quantum Information and Quantum
Physics, University of Science and Technology of China, Hefei, Anhui
230026, China

\end{affiliations}

\begin{addendum}

\item[Correspondence] Xingxing Li (lixx@ustc.edu.cn) 
\& Jinlong Yang (jlyang@ustc.edu.cn) \\

\end{addendum}
\begin{abstract}
Searching ferromagnetic semiconductor materials with electrically controllable spin polarization is a long-term challenge for spintronics. Bipolar magnetic semiconductors (BMS), with valence and conduction band edges fully spin-polarized in different spin directions, show great promise in this aspect because the carrier’s spin polarization direction can be easily tuned by voltage gate. Here, we propose a standard high-throughput computational screening scheme for searching BMS materials. The application of this scheme to the Materials Project database gives 11 intrinsic BMS materials (1 experimental and 10 theoretical) from nearly $\sim$40000 structures. Among them, a room temperature BMS \chem{Li_2V_3TeO_8 } (mp-771246) is discovered with a Curie temperature of 478 K. Moreover, the BMS feature can be maintained well when cutting the bulk Li2V3TeO8 into (001) nanofilms for realistic applications. This work provides a feasible solution for discovering novel intrinsic BMS materials from various crystal structure databases, paving the way for realizing electric-field controlled spintronic devices.

\end{abstract}

\section*{Introduction}

In the era of big data, information transmission, processing and storage are under spotlight in the research field. 
The development and utilization of the freedom of spin of electrons in this new era is the epitome of the so-called spintronics.\cite{Fert2008,Dieny2020,Wolf2001,Hirohata2020} Its outstanding features, such as faster data processing speed, higher circuit integration, lower energy consumption, and non-volatility, making it particularly advantageous in information transmission and storage.\cite{Ohno1999}  As a foundation of spintronics, functional spintronics materials such as half metallic ferromagnets (HMF), half semiconductors (HSC) and spin gapless semiconductors (SGS) have been widely investigated.\cite{Li2019,C6NR01333C,Kan2013,Zhou2011} At the same time, we have previously proposed a new conceptual material, namely, bipolar magnetic semiconductor (BMS),\cite{Li2012} which is considered to be a promising spintronics material with tunable feature. Specifically, thanks to its special band structure, the current passing through this kind of material can be totally spin polarized on the one hand, and the spin polarization direction of the current can be directly tuned by simply applying gate voltage on the other hand. 

Considering that BMS is deemed as an ideal material for electric control of spin polarization, it has become a hot spot in the field of physical chemistry in recent years. Although BMS has broad application prospects and some BMS materials have been proposed,\cite{Khalifeh2021,Zhang2014c,Zha2018,Li2013h,Zhang2021,Yuan2013,Li2016e,Li2014n} there is still a long way to go from theoretical study to experiment implementation. The primary obstacle is that the existing BMS materials are problematic in reality in the following aspects: 1) most theoretically designed BMS materials are extrinsic,\cite{Zhang2016,C9NR07793F} i.e. based on chemical or physical modifications, which makes the experimental synthesizing difficult 2) magnetic orders are only stable at very low temperatures,\cite{D0MH00183J} making it unfeasible for room temperature devices. In addition, the experimental available BMS candidates are far from sufficient on account of the low efficiency of traditional trial and error method. Therefore, it is necessary to unveil more potential BMS materials with the help of new technical means, thereby paving the way for the synthesis and application of spintronic devices. 

In recent years, with the proposal of the "Material Genome Project",\cite{NIST2011,Jain2013} high-throughput screening technology has been widely used in the field of material design, and a series of important research results have been achieved. For example, Ceder et al.\cite{Urban2016} successfully predicted one high-energy density lithium battery material\chem{LiCo_{0.5}Zr_{0.5}O_2} through high-throughput screening within the \chem{LiAl_{0.5}B_{0.5}O_2} compound structure space, which has been proved by experiments; Yan et al.,\cite{Yan2017} by integrating high-throughput theory and experimental technology, discovered 8 ternary vanadates that can be used for photocatalysis and revealed the significance of \chem{VO_4} unit in the structure. Recently, Chen et al.\cite{Chen2019} screened 40,000 compounds from the Materials Project and discovered a ferromagnetic half semiconductor (\chem{In_2MnO_7}) with a Curie temperature about 130 K. In view of the abundant achievements of high-throughput screening, we believe that it should also be adapted to search for BMS materials more effectively. 

In this work, we propose a standard high-throughput computational screening scheme for exploring BMS materials from nowadays crystal structure databases based on six filters, i.e. initialization filter, magnetic filter, stability filter, band gap filter, doping filter and refinement filters. The validity of the proposed scheme is confirmed by its successful application to the { Materials Project (MP)\cite{Ong_2015,Ong2013b} database, which contains about one hundred thousand experimentally synthesized and also theoretically predicted inorganic compounds with calculated electronic, magnetic, elastic and piezoelectric properties.}  After large scale computational screening, we obtain 11 candidate BMS materials for spintronics application, among which one BMS material with room temperature ferromagnetism has been predicted.

\section*{Results}

\begin{figure*}[ht!]
\centering
\includegraphics[width=1.0\textwidth]{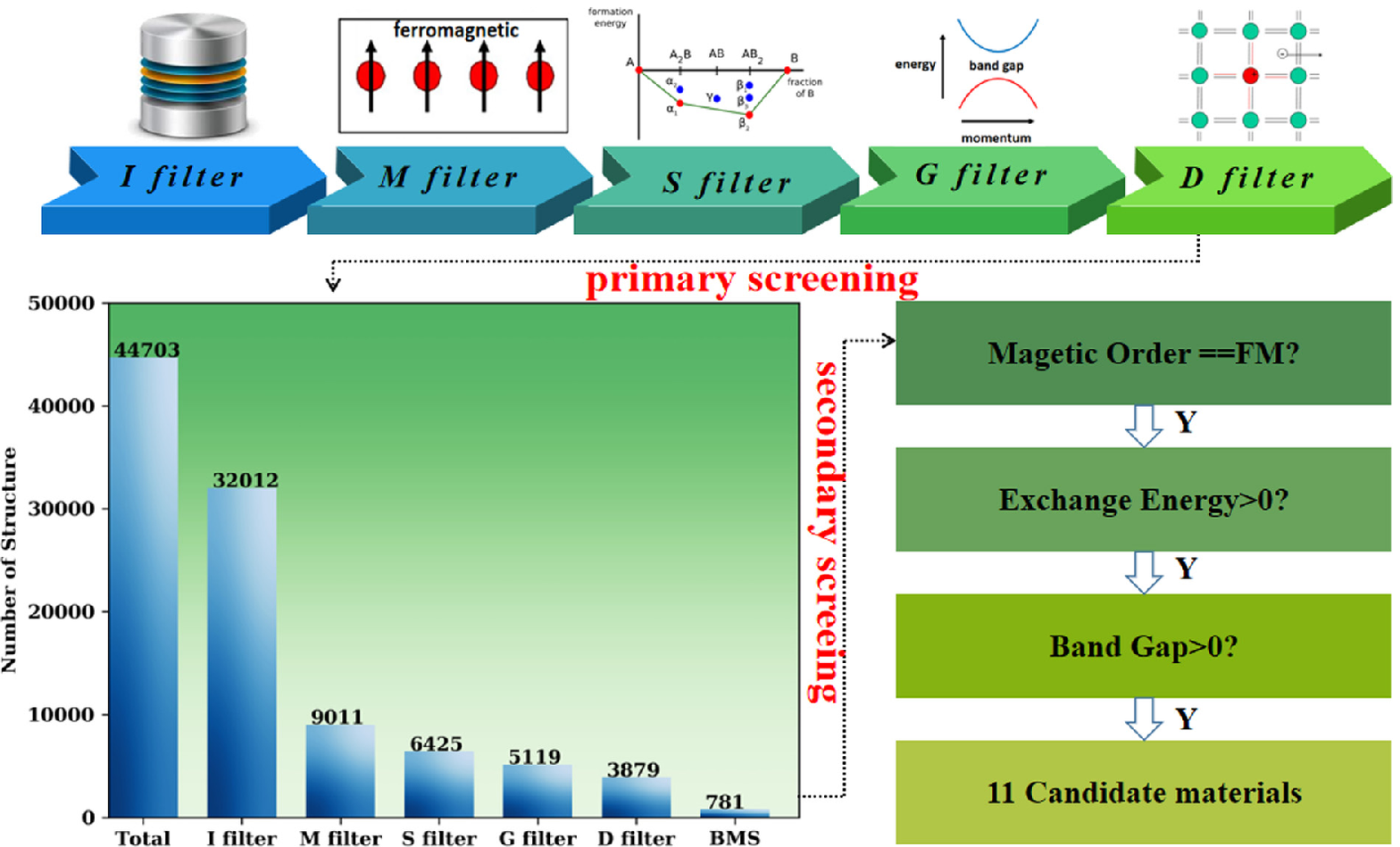}
\caption{Up panel: schematic diagram of the primary descriptors for high-throughput screening process, including Initialization filter, magnetic filter, stability filter, band gap filter and doping filter. lower-left panel: Statistic chart of the number of candidate structures in different screening stages. lower-right panel: Secondary descriptors for BMS screening.}
\label{fig:1}
\end{figure*}

As is shown in Figure \ref{fig:1}, the high-throughput screening process of BMS materials includes two groups of descriptors: primary descriptors (up panel in Figure \ref{fig:1}) and secondary descriptors (lower right panel in Figure \ref{fig:1}). The primary descriptors consist of five filters, including Initialization (I) filter, magnetic (M) filter, stability (S) filter, band gap (G) filter and doping (D) filter. The function of five filters are described as follows:
(1) \textbf{I filter}. This filter is used to build the repository for screening.  Considering the requirements of magnetic semiconductors, we select all structures with one or more elements in the set of \{ V, Cr, Mn, Fe, Co, Ni\}, where the metal and alloy structures are excluded. 
(2) \textbf{M filter}. According to band theory, one spin-polarized band has one electron. The fractional magnetic moment must have fractional occupation, which corresponds to metals. Therefore, the M filter selects the structures that only has non-zero integer magnetic moment, which, to some extent, can guarantee that the candidates are magnetic semiconductor materials. At the same time, to reduce the computational complexity, we only select the structures with ferromagnetic (FM) order.
(3) \textbf{S filter}. For stability of a compound,  two important factors are considered here, including formation energy ($E_{f}$) and the energy above convex hull ($E_{abh}$).
The formation energy\cite{Wang2020} of an compound is the energy required to produce the system from the most stable crystal structures of the individual components, which is defined as
	\begin{equation}
	\label{eq4}
	E_{f}=\frac{E(\chem{A_xB_y})-xE(\chem{A})-yE(\chem{B})}{x+y}
	\end{equation} 
where $E(\chem{A_xB_y})$ is the total energy of the material $\chem{A_xB_y}$, and $E(\chem{A})$ and  $E(\chem{B})$ are the average energies of the elements \chem{A} and \chem{B} in their stable crystal at 0K. For $E_{abh}$, it measures the energy for a material to decompose into the set of the most stable materials. A positive $E_{abh}$ indicates that this material is unstable with respect to such decomposition. A zero $E_{abh}$ indicates that this is the most stable material at its composition.\cite{Wang2020,Haastrup2018} Generally speaking, those compounds with negative $E_{f}$ are easy to be synthesized. However, considering the error of DFT calculations, we use an appropriate standard to select the candidate structures, namely, $E_{f}$ less than 0.01 eV/atom. In addition, it is generally believed that structures with $E_{abh}$ less than 0.1 eV/atom can be synthesized experimentally.\cite{Wang2020,Haastrup2018} In short, the standard for S filter is \{$E_{f}\leq0.01$ eV/atom , $E_{abh} \leq 0.1$ eV/atom\}.
(4) \textbf{G filter}. To make sure the candidate structures are semiconductors, the band gaps should be positive values. As we all know, the PBE-based band gaps value of Materials project database are underestimated. To ensure every potential BMS candidate structures to be included, the structures with band gap greater than 0.01 eV are all passed to next filter.
(5) \textbf{D filter}. Due to the special electronic structure of BMS, the magnetic moment will increase or decrease at the same time when it is doped with electrons or holes. Therefore, we use it as an important filter to screen BMS materials. In this work, the candidate structures are doped by 0.1 electron and 0.1 hole to calculate its magnetic moment.

According to the statistic chart of the number of candidate structures in primary screening stages (see in lower-left panel of Figure \ref{fig:1}), it can be found that the number of candidate structures are quickly reduced to acceptable number, specifically speaking, 781 compounds. Assuming that the data quality of Materials Project is perfect, all 781 compounds should be BMS materials. However, according to our test, it shows that some of the magnetic ground state of compounds are wrongly predicted, especially those compounds with \chem{Co} elements, which unusually has two spin states: high spin and low spin.\cite{Kremer1982} For example, the total energy and magnetic moment of entry mp-1174644  (\chem{Li_4MnCo_2O_7}) in MP database is -5.8252 eV/atom and 11.0 $\mu_B$, respectively. However, our further calculation shows that the \chem{Co} of this system should be high spin. The real total energy and magnetic moment should be -5.8466 eV/atom and 12.0 $\mu_B$, respectively. This validated results indicate that our screening in this work may miss some BMS candidate materials in MP database, which cannot be solved before the MP database is updated. At the same time, we cannot guarantee all of 781 candidate structures are technically BMS materials. To tackle this problem, we further use the secondary descriptors to refine our results.

From the lower-right panel of Figure \ref{fig:1}, it can be found that there are three steps for the secondary descriptors.\textbf{ Magnetic Order}: the magnetic ground state of 781 candidate structures are carefully investigated to search for those structures with FM order. To be specific, firstly, the FM order of all candidates are calculated to find the magnetic atom and corresponding energy $E_{FM}$. Then the antiferromagnetic (AFM) orders are set according to the number of magnetic atoms. Theoretically, the magnetic ground state calculation is a tedious task that needs to consider all combinations of magnetic order and symmetry of structure.\cite{Horton2019} To simplify the calculation, we use the following scheme. Specifically, if the number of magnetic atoms equals 2, 3, 4, 5 and 6, the number of AFM orders are 1, 2, 3, $\frac{C_5^2}{2}$ and $\frac{C_6^3}{2}$, respectively. The corresponding energies are labeled as ${E_{\{AFM\}-U}}$. If the number magnetic atoms equals 1 or is larger than 6, then we build three types of supercell ($2\times1\times1$,$1\times2\times1$ and $1\times1\times2$) to decide the magnetic ground state, where the FM order is applied for unitcell and AFM order is used between unitcells. The corresponding energies are labeled as ${E_{\{AFM\}-S}}$. Besides, considering the high spin and low spin of \chem{Co} element, all of structures with \chem{Co} element adopt 1.3 $\mu_B$ and 5.5$\mu_B$ as initial magnetic moment, which can be assigned by label "MAGMON" in the VASP input file. According to energy difference between $E_{FM}$ and ${E_{\{AFM\}-U}}$ (or ${E_{\{AFM\}-S}}/2$), the magnetic ground state is then decided.
\textbf{Exchange Energy}: The Curie temperature is closely related to magnetic exchange energy ($E_{ex}$). According to the above label, the $E_{ex}$ is defined as: $E_{ex}$=max(${E_{\{AFM\}-S}}$/2-$E_{FM}$) or $E_{ex}$=max(${E_{\{AFM\}-U}}$-$E_{FM}$). 
\textbf{Band Gap}: According to the refined magnetic ground state, the band gap of selected structures are recalculated by HSE06 functional to identify those structures with positive band gap and BMS feature.

\begin{figure*}[!htb]
\centering
\includegraphics[width=1.0\textwidth]{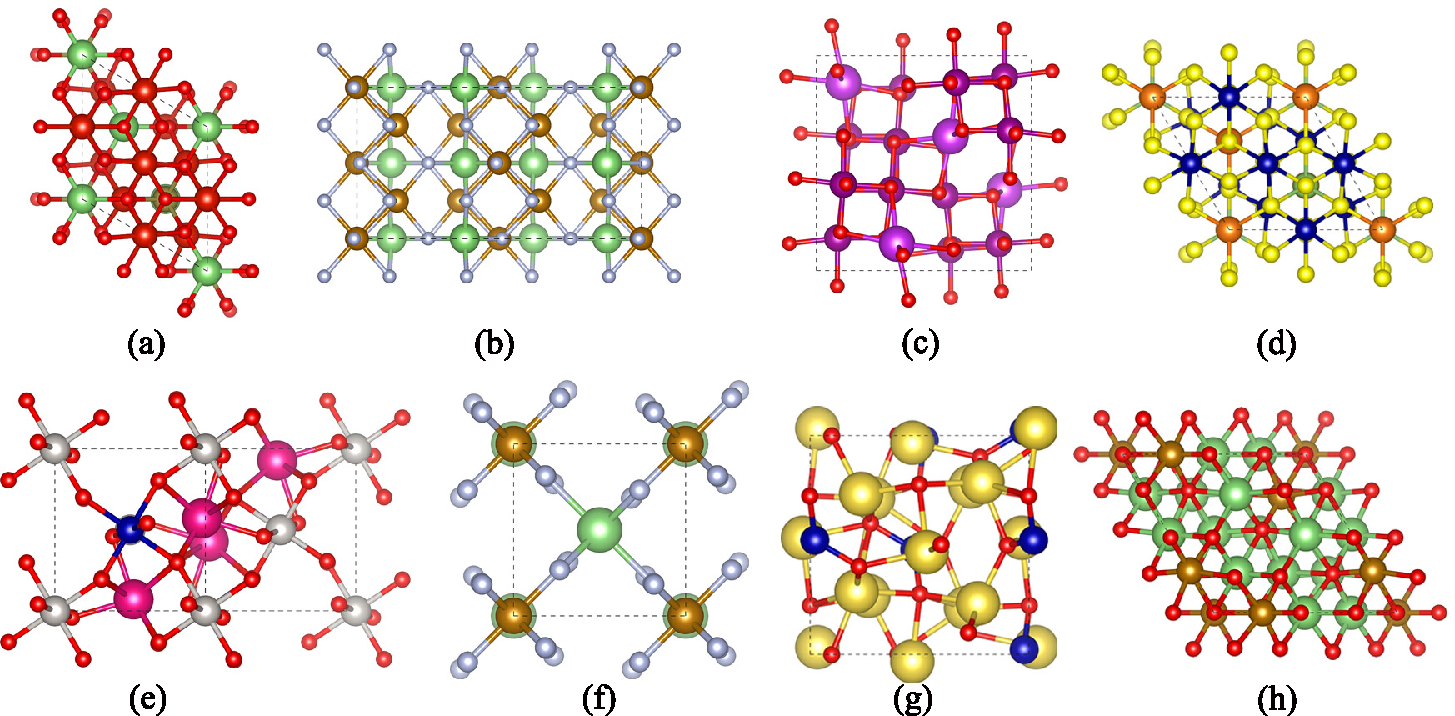}
\caption{ (a)-(h) Crystal structure of \chem{Li_2V_3TeO_8 } (mp-771246),  \chem{Li_2Fe_3F_8} (mp-1177989),  \chem{Mn_3BiO_8} (mp-773037),  \chem{Mg_2Cr_3GaS_8} (mp-1247148), \chem{Sm_2CoPtO_6 } (mp-1208920), \chem{ LiFeF_6} (mp-1222351), \chem{Na_3CoO_3} (mp-755811) and \chem{Li_3FeO_4} (mp-849528),  respectively}
\label{fig:2}
\end{figure*}

Based on the screening scheme as described in Figure \ref{fig:1}, we finally obtain 11 candidate BMS materials for spintronics, among which 1 structure (\chem{CoPtF_6}, MP-ID: mp-556492, ICSD: 37447 ) is experimentally synthesized and 10 structures are theoretically predicted. Properties of obtained BMS materials are listed in Table \ref{table:1}, which is sorted by the exchange energy per magnetic atom. According to the statistic, the number of entries for \chem{V}-, \chem{Cr}-, \chem{Mn}-, \chem{Fe}-, \chem{Co}- and \chem{Ni}-based BMS are 1, 1, 2, 3, 3 and 1, respectively. Although 10 of 11 candidate BMS candidates are theoretically predicted structures, all of them have negative $E_{f}$ and zero or small positive $E_{abh}$, which indicates that these structures may be synthesized by experiments in future. It is worth noting that the $E_{ex}$ of mp-771246 is larger than 100 meV/atom, indicating it may have a high Curie temperature and be promising in spintronics application.  For the rest structures, all of $E_{ex}$ are less than 100 meV/atom and show a general tendency: the larger magnetic atom distance is, the smaller $E_{ex}$ is. In order to have a clear understanding, we present the crystal structure of first 8 materials (The structures of left BMSs are supplied in the SI Figure S1) in Figure \ref{fig:2} due to their relatively large $E_{ex}$ .
It can be found that, all of structures have high symmetry and most of the unit cell of structures have at least 2 magnetic atoms.

In addition, according to our screening results, the energy of FM state and AFM are nearly degenerate for 6 entries, including \chem{Co(HO)_2} (mp-24105),  \chem{CaMg_{14}CoO_{16}} (mp-1036443),  \chem{Dy_2CoTe_2(SO_7)_2} (mp-1190177),  \chem{MnH_8(NF_3)_2} (mp-759690), \chem{K_2MnF_6} (mp-560127) and \chem{Li_4NiSn_3(PO_4)_4} (mp-776070) (See SI Table S1). Previous study shows that strain or doping can be an effective way to tune AFM state to FM state.\cite{Li2016e} As a test case, we use 2\% strain for mp-759690(\chem{MnH_8(NF_3)_2}) to check its strain response. The results show that the ground state magnetic order of \chem{MnH_8(NF_3)_2} under this small strain is FM state and also present BMS feature (The corresponding DOS is supplied in the SI Figure S2). 

\begin{figure*}[!htp]
\centering
\includegraphics[width=1.0\textwidth]{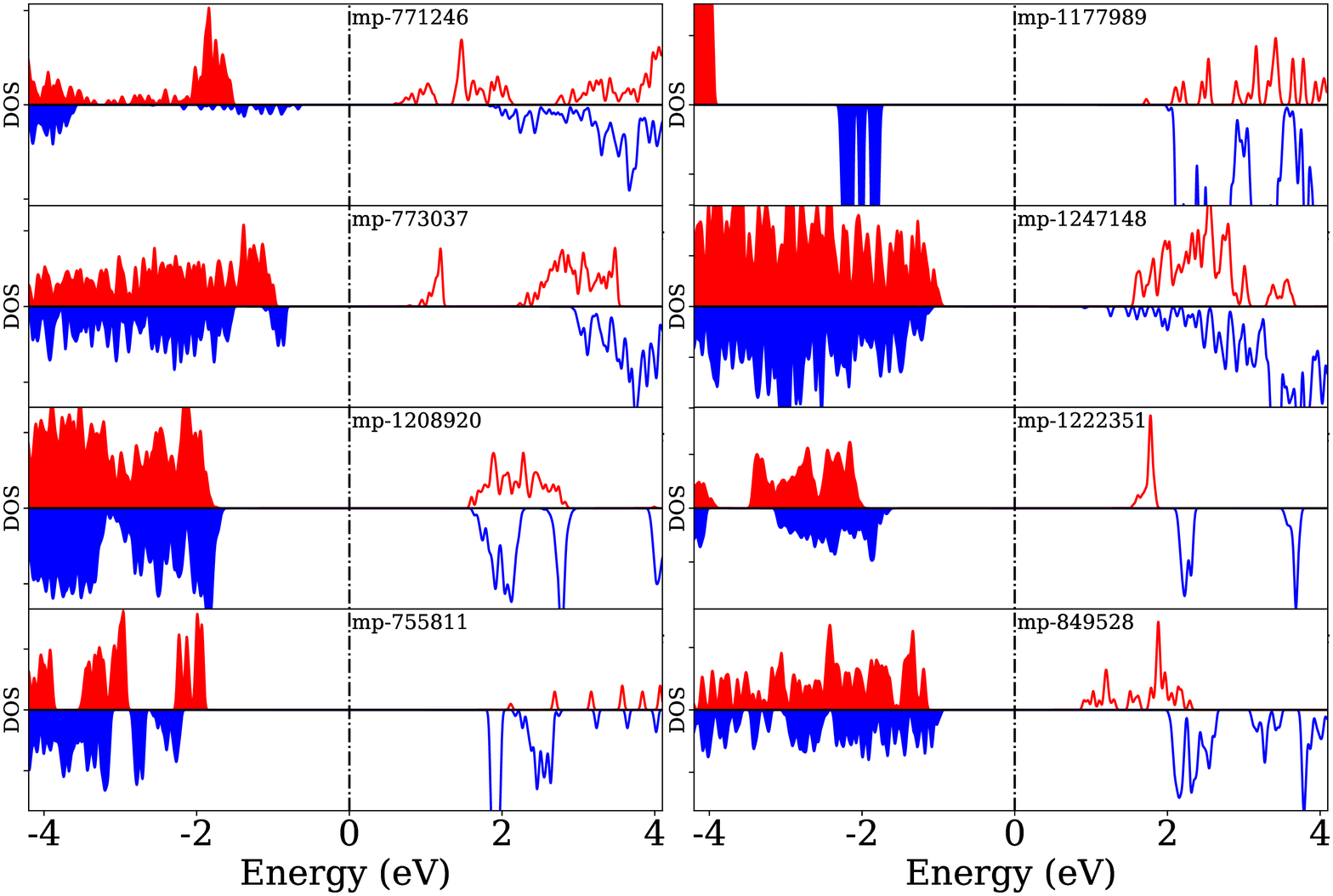}
\caption{HSE06 level density of states for \chem{Li_2V_3TeO_8 } (mp-771246),  \chem{Li_2Fe_3F_8} (mp-1177989),  \chem{Mn_3BiO_8} (mp-773037),
 \chem{Mg_2Cr_3GaS_8} (mp-1247148), \chem{Sm_2CoPtO_6 } (mp-1208920), \chem{ LiFeF_6} (mp-1222351), \chem{Na_3CoO_3} (mp-755811 ) and \chem{Li_3FeO_4} (mp-849528 ),  respectively}
\label{fig:3}
\end{figure*}

Then the basic electronic structure properties of obtained BMSs are investigated. In Figure \ref{fig:3} we plotted the DOS for first 8 structures by using HSE06-based first-principles calculations (The corresponding HSE06-based DOS for left entries are supplied in the SI Figure S3). We can find that all of 8 compounds possess completely spin polarized DOS with opposite spin orientations around the Fermi level. Due to the unique electronic structure, the spin direction of these BMS systems can be easily tuned by the electrical gating technique. A positive/negative gate voltage will inject electrons/holes into the BMS system, which causes a controllable half-metallic conducting behaviour.\cite{Li2012} 

\begin{figure}[!htp]
\centering
\includegraphics[width=1.0\textwidth]{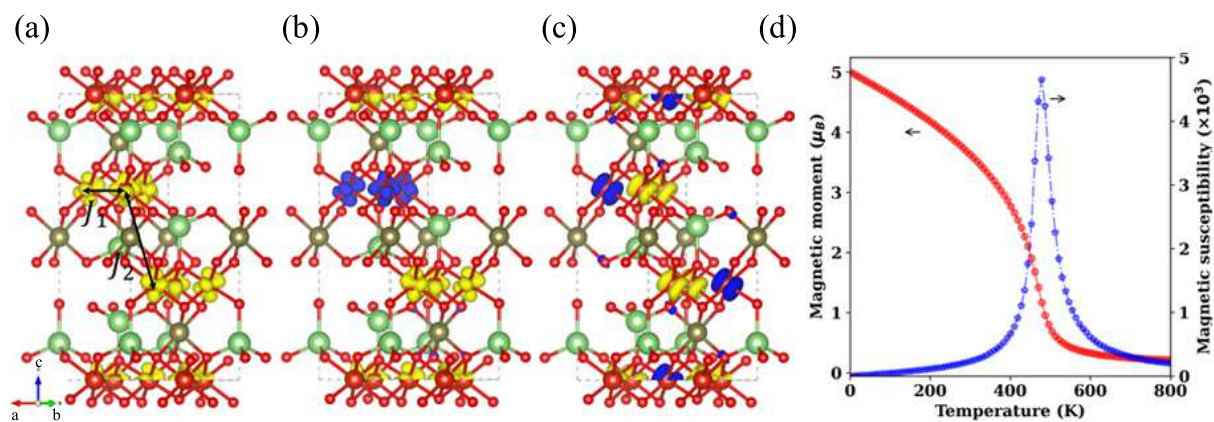}
\caption{Isosurface of spin density of (a) ferromagnetic state and (b)-(c) antiferromagnetic state of \chem{Li_2V_3TeO_8 } with an isovalue of 0.031 eV/$\AA^3$. Yellow and blue indicate the positive and negative values, respectively. (d) The evolution of spin magnetic moment  (red circle)  per unit cell and magnetic susceptibility (green pentagon) with respect to temperature  for \chem{Li_2V_3TeO_8 } (mp-771246). }
\label{fig:4}  
\end{figure}

To estimate the Curie temperatures of obtained BMS materials, three methods are available, including mean-filed approximation (MFA), random-phase approximation (RPA) and Monte Carlo simulation (MC).\cite{PhysRevB.73.214412} Compared with RPA and MC, MFA usually overestimates the Curie temperature. In this work, we use MC method based on the classical Heisenberg Hamiltonian to estimate the Curie temperature: 
	\begin{equation}
	\label{eq2}
	H=-\sum_{i,j}{J_{i,j}S_{i}S_{j}}
	\end{equation} 
where $J_{i,j}$ is the exchange parameter and $S$ is the spin of magnetic atoms. Here, we consider two exchange parameters $J_{1}$ and $J_{2}$ (Labeled in Figure \ref{fig:4}(a)), which represent the intralayer and interlayer nearest-neighbor exchange parameter respectively. The spin density of FM and AFM states are shown in Figure \ref{fig:4}(a)-(c), where the local magnetic moments are all contributed by \chem{V} atoms. For MC simulation, the supercell for \chem{Li_2V_3TeO_8 } is set to be $8\times8\times8$.  
According to the $E_{ex}$ in Table \ref{table:1}, we deduce that the $J_{i,j}$ of \chem{Li_2V_3TeO_8} is 33.3 meV and  15.7 meV. Then the simulated spin magnetic moment and susceptibility as a function of temperature are plotted in Figure \ref{fig:4}(d). For \chem{Li_2V_3TeO_8}, all the spins of V atoms point in the same direction at 0K, forming a strict FM order, while the magnetic moment decreases rapidly when heated. The critical point for FM to paramagnetic transition occurs at about 478 K, as indicated by an abrupt increase in the magnetic susceptibility curve. 

As discussed above, the bulk \chem{Li_2V_3TeO_8} is predicted to be a potential candidate for BMS. However, to design modern nano-sized devices, a thin film (slab) structure is required. Due to the surface states and quantum-size effect, the properties of two-dimensional (2D) slab structures are generally different from their bulk counterpart and depend on the slab’s thickness. Therefore, it is necessary to answer the following  questions: Can the 2D slab of \chem{Li_2V_3TeO_8} maintain the BMS feature? What is the smallest thickness required for maintaining stable BMS feature? For simplicity, we  here only consider the (001) oriented slabs. According to the symmetry of \chem{Li_2V_3TeO_8}, 5 slabs with different surface terminations are selected, namely, \chem{O}1-, \chem{O}2-, \chem{Te}-, \chem{V}- and \chem{Li}-termination (See Figure S4). The calculation shows that the \chem{O}1-termination has the lowest formation energy. Then the electronic properties of \chem{O}1-termination with different number of \chem{V} atom layers are tested. 
Considering that FM order will finally be the ground state as the number of \chem{V} atom layers increases, here we only calculate the DOS of \chem{O}1-terminated slabs with FM order to reduce the workload. According to DOS in Figure S5, it can be found that as the number of \chem{V} atom layers increases, the electronic structure gradually converges. Specifically, when the number of \chem{V} atom layers is greater than or equal to 6, the total magnetic moment will increase by 5 $\mu_B$ for each additional \chem{V} atom layer. At the same time, when the number of \chem{V} atom layers reaches to 11, the slab structure shows BMS feature and the structure with 12 \chem{V} atom layers also has the same feature and a similar DOS. That is to say, a minimum number of 11 \chem{V} atom layers (corresponding to 5.34 nm) are required for spintronics application. The detailed properties of 2D \chem{Li_2V_3TeO_8} slabs with different thicknesses are summarized in Table S2. 

Despite its potential application in spintronics devices design, the \chem{Li_2V_3TeO_8} is still not synthesized yet. However, according to the query results from MP database (see Table S3), it can be found that the formation energy and energy above hull are all lower than that of experimental phases \chem{LiVTeO_5 } (ICSD: 21012 ) and \chem{LiV_3(TeO_6)_2 } (ICSD: 249325)  when the stable phases \chem{LiTe_3} (ICSD: 935), \chem{VTe_2} (ICSD: 38369) and \chem{TeO_2} (ICSD: 26844 and 30222) are selected as references. At the same time, the theoretically proposed  reaction of $2 \mathrm{LiTe}_{3}+3 \mathrm{VTe}_{2}+15 \mathrm{O}_{2} \rightarrow 11 \mathrm{TeO}_{2}+\mathrm{Li}_{2}\mathrm{V}_{3}\mathrm{TeO}_{8}$ has the largest reaction enthalpy change (exothermic reaction) compared with \chem{LiVTeO_5 } and  \chem{LiV_3(TeO_6)_2 }. Therefore, \chem{Li_2V_3TeO_8} would have possibility to be synthesized experimentally. 

\section*{Discussion}
In summary, based on initialization filter, magnetic filter, stability filter, band gap filter, doping filter and refinement filters, we propose a standard high-throughput computational screening scheme for exploring an important class of ferromagnetic semiconductors with electrically controllable spin polarization, namely BMS. Compared with the traditional trial and error method, the present scheme is direct and efficient. By using present scheme, a total number of 11 BMS materials are obtained via screening a comprehensive quantum materials repository containing 44703 magnetic compounds. Among them, \chem{Li_2V_3TeO_8} is predicted to be a room temperature BMS with a Curie temperature about 478 K. Meanwhile, the corresponding slab structure of \chem{Li_2V_3TeO_8} with 11 or more \chem{V} atom layers can maintain the bulk's BMS feature well. Besides, the low formation energy and energy above convex hull make the experimental synthesis of \chem{Li_2V_3TeO_8} feasible.

However, some drawbacks also exist for current scheme. For example, the DFT calculation is very expensive, which limits us to search for BMS materials with larger unitcell. In addition, the current MP database is not adequately qualified, which will result in the missing of some potential BMS candidates. To solve these problems, highly qualified database is needed. At the same time, based on existing BMS materials to build structure-properties relationship, then use machine learning method to predict new BMS is a feasible way. We will leave these problems in our future work and also wish our current work can guide experiments and theories for developing BMS materials.

\begin{methods}

The first-principles calculations of electronic structure are conducted by using  the Vienna ab initio simulation package (VASP) software package.\cite{Kresse1996,KRESSE199615} The generalized gradient approximation (GGA) of Perdew, Burke, and Ernzerhof (PBE) exchange correlation functional \cite{perdew1996generalized} with collinear spin polarization is employed. The plane wave basis set is used to describe the valence electrons with the cutoff set to 520 eV. The convergence criteria for electronic SCF iterations and ionic step iterations were set to be $1.0\times10^{-6}$ eV and $0.5\times10^{-3}$ eV$\AA^{-1}$, respectively. The reciprocal space grid was set to 8000 according to Pymatgen method.\cite{Ong2013b} To better describe onsite Coulomb repulsion among \textit{d} or \textit{f} electrons, we adopt the GGA+U scheme\cite{PhysRevB.52.R5467} during the screening stage, where the effective onsite coulomb interaction parameter ($U$) and exchange interaction parameter ($J$) for different structures are set according to Pymatgen default values. To fix semi-local PBE calculations of electronic structures, the density of states (DOS) of final candidate materials are calculated based on screened hybrid HSE06 functional\cite{heyd2003hybrid, ge2006erratum} with 20\% Hartree-Fock exchange.

In order to ensure that the large scale BMS screening tasks running on the HPC cluster in an efficient, stable, and automated manner, we use dpdispather (see in Code Availability) to create, manage and collect tasks. To generate input files of static and density of states calculation and to analyse output files, Pymatgen and maptool (see in Code Availability) are used. The visualization of crystal structure is implemented by VESTA software.\cite{Momma2008}

\end{methods}

\section*{Acknowledgements}
\begin{addendum}
 \item[Author Contributions]  Haidi Wang, XingXing Li and Jinlong Yang conceived the idea. Haidi Wang and Qingqing Feng performed the calculations. All authors helped to write, modify and analyze this manuscript. 
 \item[Funding] This work is financially supported by the Fundamental Research Funds for the Central Universities, by the National Natural Science Foundation of China (Grant No. 21688102), by the National Key Research \& Development Program of China (Grant No. 2016YFA0200604), by Anhui Initiative in Quantum Information Technologies (Grant No. AHY090400), by the Youth Innovation Promotion Association CAS (2019441), and by USTC Research Funds of the Double First-Class Initiative (YD2060002011). The authors thank the Supercomputer platform of USTC, Beijing and HFUT for the computational resources.  
 \item[Conflicts of Interest ] The authors declare that there are no competing interests. 
 \item[Data Availability] The authors confirm that the data supporting the findings of this study are available within the article and its supplementary materials. All of the structures and related information can be downloaded online (https://gitee.com/haidi-hfut/bms).
 \item [Code Availability] The dpdispatcher can be found at (https://github.com/deepmodeling/dpdispatcher).
 The maptool can be found at (https://github.com/haidi-ustc/maptool).
 \end{addendum}

\section*{Supplementary Materials} Table for properties of entries mp-24105, mp-1036443, mp-1190177, mp-759690, mp-560127 and mp-776070;  Crystal structure and density of state for \chem{Li_4Ni_7(OF_7)_2 } (mp-867641), \chem{ CoPtF_6 }(mp-556492 ) and \chem{ Li_2MnF_6}(mp-754966); density of state for \chem{MnH_8(NF_3)_2} under strain;  density of state for \chem{Li_4Ni_7(OF_7)_2 } (mp-867641), \chem{ CoPtF_6 }(mp-556492 ) and \chem{ Li_2MnF_6}(mp-754966); table for properties of \chem{Li_2V_3TeO_8 } of (001) oriented \chem{O}1-termination slabs with different number of \chem{V} atom layers; (001) oriented slabs of \chem{Li_2V_3TeO_8} with different surface terminations; density of state of \chem{Li_2V_3TeO_8 } of (001) oriented \chem{O}1-termination slabs; table for properties of Li-V-Te-O quaternary compounds \chem{Li_2V_3TeO_8 }, \chem{LiVTeO_5 } and \chem{LiV_3(TeO_6)_2 }.

\begin{addendum}
\item[References]

\end{addendum}

\footnotesize{
\bibliography{refs}
}

\clearpage
\begin{addendum}

\item[Figures and Table]

\end{addendum}

\begin{table*}[!htb]
	\begin{center}
		\caption{\label{table:1} Properties of obtained compounds for BMS materials based PBE level simulation: Materials Project ID (MP-ID), formula, formation energy ($E_{f}$,eV/atom), energy above hull ($E_{abh}$,eV/atom), magnetic moment per formula unit cell ($m$,$\mu_B$/f.u.), minimum distance between magnetic atoms ($d_{min}$,$\AA$), number of magnetic atom ($N_m$), exchange energy per magnetic atom ($E_{ex}$,meV/atom) and space group symmetry. The spin-flip gap in valence band ($\Delta_1$, eV), conduction band ($\Delta_2$, eV) and the band gap ($\Delta_3$, eV) }
		\setlength{\tabcolsep}{1mm}{
\begin{tabular}{llcccccccccc}
\hline
MP-ID &        formula &     $E_{f}$ &  $E_{abh}$ &  $m$ & $d_{min}$ &  $N_m$ & $E_{ex}$ & symmetry & $\Delta_1$ & $\Delta_2$ & $\Delta_3$ \\
\hline
mp-771246  &      \chem{Li_2V_3TeO_8 }&  -2.270 &  0.045 &      6 &    3.021 &               3 &   180 &      R-3m &   0.855 &   1.259 &   0.254 \\
mp-1177989 &       \chem{Li_2Fe_3F_8} &  -2.863 &  0.034 &     13 &    3.168 &               6 &   71 &      Cmce &   1.242 &   0.414 &   3.124 \\
mp-773037  &         \chem{Mn_3BiO_8} &  -1.673 &  0.060 &      8 &    2.898 &              12 &   49 &    $P4_332$ &   0.068 &   2.179 &   0.613 \\
mp-1247148 &     \chem{Mg_2Cr_3GaS_8} &  -1.252 &  0.041 &      9 &    3.622 &               3 &   21 &      R-3m &   0.450 &   0.507 &   0.366 \\
mp-1208920 &       \chem{Sm_2CoPtO_6 }&  -2.566 &  0.005 &      4 &    5.427 &               2 &   19 &    $P2_1/c$ &   0.127 &   0.025 &   1.627 \\
mp-1222351 &          \chem{ LiFeF_6} &  -2.265 &  0.000 &      4 &    4.738 &               2 &   3 &    $P4_2nm$ &   0.121 &   1.365 &   1.124 \\
mp-755811  &         \chem{Na_3CoO_3} &  -1.562 &  0.038 &      5 &    5.101 &               4 &   3 &     $P2_13$ &   0.663 &   0.340 &   1.803 \\
mp-849528  &         \chem{Li_3FeO_4} &  -1.872 &  0.015 &      3 &    3.082 &               4 &   2 &     I-43m &   0.115 &   1.769 &   0.231 \\
mp-867641  &  \chem{Li_4Ni_7(OF_7)_2 }&  -2.264 &  0.062 &     15 &    2.975 &               7 &   2 &      C2/m &   0.440 &   0.308 &   2.815 \\
mp-556492  &          \chem{ CoPtF_6 }&  -1.997 &  0.000 &      4 &    5.192 &               1 &   2 &       R-3 &   0.040 &   0.020 &   1.841 \\
mp-754966  &        \chem{ Li_2MnF_6} &  -2.888 &  0.000 &      4 &    4.667 &               2 &   1 &  $P4_2/mnm$ &   0.073 &   1.418 &   2.617 \\
\hline
\end{tabular}
}
	\end{center}
\end{table*}

\end{document}